\documentclass[pre,superscriptaddress,showpacs,nofootinbib,floatfix,preprint]{revtex4}
\usepackage{graphicx}
\usepackage{dcolumn}
\usepackage{bm}
\usepackage{epsfig}
\usepackage{booktabs,topcapt}
\begin{document}

\preprint{PREPRINT}

\title{Effects of the attractive interactions in the thermodynamic, dynamic
and structural anomalies of a two length scale potential}

\author{Jonathas Nunes da Silva}
\affiliation{Instituto de F\'{\i}sica, Universidade Federal do Rio Grande do
Sul, Caixa Postal 15051, 91501-970, Porto Alegre, RS, Brazil}
\email{jonathas@if.ufrgs.br}

\author{Evy Salcedo}
\affiliation{Departamento de F\'{\i}sica,
Universidade Federal de Santa Catarina, Florian\'opolis, SC, 
88010-970, Brazil}
\email{esalcedo@fsc.ufsc.br}

\author{Alan Barros de Oliveira}
\affiliation{Departamento de F\'{i}sica, Universidade Federal de Ouro Preto,
Ouro Preto, MG, 35400-000, Brazil}
\email{oliveira@iceb.ufop.br}

\author{Marcia C. Barbosa}
\affiliation{Instituto de F\'{\i}sica, Universidade Federal do Rio Grande do
Sul, Caixa Postal 15051, 91501-970, Porto Alegre, RS, Brazil}
\email{marcia.barbosa@ufrgs.br}

\date{\today}
\begin{abstract}
Using molecular dynamic simulations we study a system
of particles interacting through a  continuous core-softened 
potentials consisting of a hard core, a shoulder at closest distances and an 
attractive well at further distance.  We obtain the pressure-temperature 
phase diagram of  of this system for various depths of the  tunable attractive 
well. Since this is a two length scales potential, density, diffusion
and structural anomalies are expected. 
We show that the effect of increasing the attractive interaction
between the molecules is to shrink the region in pressure in which
the density and the diffusion anomalies are present. If the 
attractive forces are too strong, particle
will be predominantly in one of the two length scales and
no density of diffusion anomaly is observed.  
The structural anomalous region is present
for all the cases.

\end{abstract}

\maketitle

\section{Introduction} 
\label{intro}

The phase behavior of single component systems as particles 
interacting via the so-called core-softened (CS) potentials
are receiving a lot of attention recently. 
These potentials exhibit a repulsive core with
a softening region with a shoulder  or 
a ramp \cite{Bu03,Sk04,He05b,He70,Ja98,Wi02,Ol06a,Al09,Fo08,Fr07a,Ol08b,Ol09}.
These  models were motivated by the aim of construct a 
simple two-body isotropic potential capable of describing
the complicated features of systems interacting via anisotropic 
potentials. This approach generates models 
analytically \cite{Zh06,Zh08,Zh09,Eg08a} and
computationally \cite{Bu03,Sk04,He05b,He70,Ja98,Wi02,Ol06a,Al09,Fo08} 
tractable still capable
to retain the qualitative features of the real complex systems.

The physical motivation behind these studies is the recently acknowledged 
possibility  that some single 
component systems  display coexistence
between two different liquid phases \cite{Fr01,Po92,Gl99}, a low
density liquid phase (LDL) and a high density liquid phase (HDL),
ending at a LDL-HDL critical point. This 
opened the 
discussion about the relation between the presence of 
two liquid phases, the existence of thermodynamic anomalies in liquids
and the form of the potential.
The case of water is probably the most intensively studied.  
A liquid where the specific volume at ambient pressure starts to increase 
when cooled below $T\approx 4 ^oC $ \cite{Wa64,An76}. Besides, in a certain
range of pressures, water also exhibits an anomalous increase of 
compressibility and specific heat upon cooling   from
 experiments~\cite{Pr87,Ha84}.
Experiments for Te, \cite{Th76} Ga, Bi,~\cite{LosAlamos} S,~\cite{Sa67,Ke83} 
and Ge$_{15}$Te$_{85}$,~\cite{Ts91}  and 
simulations for silica,~\cite{An00,Ru06b,Sh02}
silicon~\cite{Sa03} and BeF$_2$,~\cite{An00} show that these
materials present also density anomaly.

Besides the anomalies  discussed above, water has dynamic anomalies as well. 
Experiments  show that the diffusion 
constant, $D$, increases on compression at low temperature, $T$,  up to a 
maximum $D_{\rm max} (T)$ at 
$P =P_{D\mathrm{max}}(T)$. The behavior of normal liquids, with 
$D$  decreasing on 
compression, is 
restored in water only at high $P$, e.g. for 
$P > P_{D\mathrm{max}}\approx 1.1$ kbar  at $10^o$C ~\cite{An76,Pr87}.
Computational simulations for the 
Simple Point Charge/Extended (SPC/E) water model \cite{spce} 
recover the experimental results and 
show that the anomalous behavior of $D$ extends to the metastable 
liquid phase of water at negative pressures --
a region that is difficult to access for 
experiments \cite{Ne01,Ne02b,Er01}. 
In this region the diffusivity $D$ decreases 
for decreasing $p$ until it reaches 
a minimum value $D_{\rm min} (T)$ at some pressure $p_{D\mathrm{min}}(T)$, and 
the normal behavior, with $D$ increasing for
decreasing $p$, is reestablished only for 
$P < P_{D\mathrm{min}}(T)$~\cite{Ne01,Ne02b,Er01,Mi06a,Mu05}. Besides 
water, silica~\cite{Sh02, Ch06} and silicon~\cite{Mo05} also exhibit a 
diffusion anomalous region.

Acknowledging that CS potentials might engender density and diffusion
anomalies, de Oliveira \emph{et al.} \cite{Ol06a,Ol06b,Ol07,Ol08a,Ol10a,Ol10b}
proposed a simple CS model.
It has a repulsive core that exhibits a region
of softening where the slope changes drastically. This
model  exhibits density, diffusion and
structural anomalies like the anomalies present in
experiments \cite{An76,Pr87} and simulations \cite{Ne01,Ne02b,Er01} for water.
This simple system has no attraction between the 
particles and, therefore, no
liquid-gas or liquid-liquid critical points are present. Realistic
models should have attractive interactions since most
molecules attract each other either due to
van der Waals interactions or to more
sophisticated electrostatic forces.

Which effect in the pressure-temperature phase diagram
one might expect from the addition of a larger
attractive part in the potential? For one length scale potentials,
the  increase of the attractive well leads to 
an increase in the temperature of
the  liquid-gas critical point. In the case 
of the continuous two length scale
potential the same behavior might
be expected for the liquid-gas critical 
point but  it is not clear which effect the depth of the well has
 in the location in the pressure-temperature phase diagram of 
the  liquid-liquid critical point. Moreover,
it is also not clear which effect the attraction has 
in the location in the pressure-temperature phase diagram of 
the density, diffusion and structural anomalous regions.

In this paper we address these two questions
by studying  the pressure-temperature phase diagram of
a potential with a repulsive core followed by a tunable attractive well.
We check if the introduction of the attraction between particles
affects the liquid-liquid critical point and the density, diffusion
and structural anomalies.

The remaining of this paper goes as follows. In Sec. \ref{model} the
model is introduced and the methods  are presented.
Details of simulations are given Sec. \ref{simulations}.
In Sec. \ref{results} the results are discussed and, finally, 
the conclusion are made in Sec. \ref{conc}.

\section{ The Model} 
\label{model}

The model consists of a system of $N$ particles of diameter 
$\sigma$, inside a cubic box with  volume  $V$ , resulting in a 
number density $\rho = N/V$ . The interacting effective potential 
between particles is given by
\begin{equation}
U^{*}(r) = 4  \left[ \left( \frac{\sigma}{r} \right)^{12} - 
\left( \frac{\sigma}{r}\right)^{6} \right] + a\exp \left[ -
\frac{1}{c^{2}} \left( \frac{r - r_{0}}{\sigma}
\right)^{2} \right] + b\exp \left[ -
\frac{1}{d^{2}}\left( \frac{r-r_{1}}{\sigma} \right)^{2} \right] \;,
\label{eq:potential}
\end{equation}
where $U^{*}(r) = U(r)/\varepsilon$. The first term of 
Eq. (\ref{eq:potential}) is a 
Lennard-Jones potential of well depth $\varepsilon$.  The second and third
terms are Gaussians centered on radius $r = r_{0}$ and 
$r = r_{1}$, with heights $a$ and 
$b$, and widths $c$ and $d$ respectively. 
This potential can represent a whole family of
two length scales  intermolecular interactions, from a deep double
wells potential \cite{Ch96b,Ne04} to a
repulsive shoulder \cite{Ja98}, depending on the choice of the values
of the parameters.

For $b=0$ the attractive part vanishes and the potential
becomes purely repulsive. This case was previously studied
for determining the pressure-temperature phase diagram as
well as the regions where water-like anomalies occur \cite{Ol06a,Ol06b}.

\begin{figure}[htb]
  \begin{centering}
   \includegraphics[clip=true,width=8cm]{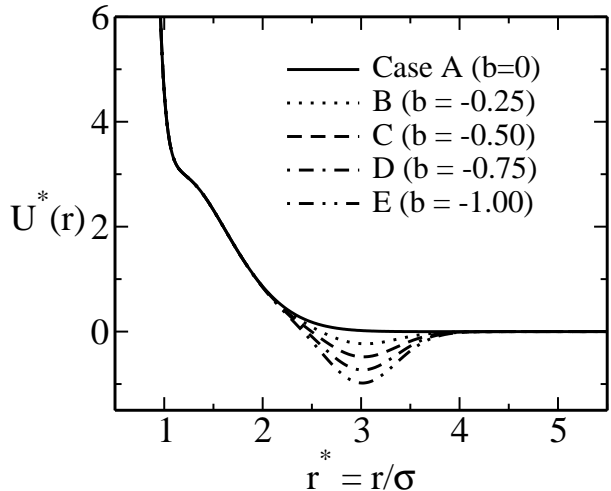} 
    \par
  \end{centering}
  \caption{Interaction  potential Eq.~(\ref{eq:potential}) with 
parameters $a = 5.0$, $r_{0}/\sigma = 0.7$, $c = 1.0$, $r_{1}/\sigma = 3.0$ 
and $d = 0.5$ for all cases. $b$ is shown in Table \ref{tab:b} 
for each case.}
  \label{fig:potential} 
\end{figure}

How the addition of an attractive part in the potential
affects the overall pressure-temperature phase diagram?
In order to answer to this question we obtain the 
pressure temperature phase diagram of the potentials
illustrated in Fig.~\ref{fig:potential} where the attractive part
is increased systematically without changing the core-softened part
of the potential.
This is done by setting the potential given by
Eq.~(\ref{eq:potential})  with fixed parameters:  $a = 5.0$, 
$r_{0}/\sigma = 0.7$, $c = 1.0$,  $r_{1}/\sigma = 3.0$,
$d = 0.5$ for the five cases studied in this work.
The parameter $b$ for each case is
shown in Table~\ref{tab:b}. 

\begin{table}[htbp]
   \centering
   \topcaption{Parameter $b$ in the potential Eq. (\ref{eq:potential}) 
for each case studied in this work. 
   The other parameters are $a = 5.0$, 
$r_{0}/\sigma = 0.7$, $c = 1.0$,  $r_{1}/\sigma = 3.0$, and
$d = 0.5$ for the five cases.} 
   \begin{tabular}{@{} lccc @{}  } 
      \toprule
     \\ \hline \hline
      
                            &&& $b$     \\ \hline \hline
      \midrule
       Case A          &&&       0    \\  \hline
       Case B         &&&  $-0.25$\\  \hline
       Case C         &&&  $-0.50$\\ \hline
       Case D         &&&  $-0.75$ \\ \hline 
      Case E         &&&  $-1.00$ \\ \hline 
\hline
     \bottomrule
   \end{tabular}
   \label{tab:b}
\end{table}

The potential shown in Fig. \ref{fig:potential} 
has two length scales within a repulsive 
shoulder followed by a attractive well.
The addition of an attractive part 
to the ramp-like format gives rise to a liquid-liquid 
first order phase transition 
and to a first order liquid-gas phase transition
ending at  critical points. The liquid-liquid phase transition
is located in the vicinity of the anomalous region.

\section{Details of Simulations} 
\label{simulations}

For the case in which $b=0$ the results shown in this paper were
adapted from Refs. \cite{Ol06a,Ol06b}. For the other cases ($b\neq0$)
the details of simulations are as follows.

The quantities of interest were obtained by $NVT$-constant molecular 
dynamics using the LAMMPS package \cite{lammps}.
$N=1372$ particles were used into a cubic  
box with periodic boundary conditions in all directions.
The interaction through particles,
Eq.~(\ref{eq:potential}), had a cutoff of $4.5\sigma$ and 
the Nose-Hoover heat-bath 
was used in order to keep fixed the temperature.

All simulations were initialized in a liquid phase previously 
equilibrated over $5\times 10^5$ steps at $T^{*} = 0.6$. 
The time step used was $0.001$ in reduced units and
the runs were carried out for a
total of $3\times10^6$ steps, dumping instantaneous 
configurations for every 2000 steps, giving then a total
of 1500 independent configurations. 
The first 200 configuration were discarded  
for equilibration purposes, thus 1300 configurations were
used for sampling averages.

Temperature, pressure,  density  and diffusion 
are shown in dimensionless units,

\begin{eqnarray}
T^{*}&\equiv & \frac{k_{B}T}{\epsilon}\; \nonumber \\
\rho^{*}&\equiv &\rho \sigma^{3} \nonumber \\
P^*&\equiv& \frac{P \sigma^{3}}{\epsilon} \nonumber \\
D^*&\equiv& \frac{D(m/\epsilon)^{1/2}}{\sigma}\; .
\end{eqnarray}

The pressure of the system is calculated by means of the the virial 
theorem,

\begin{eqnarray}
P&=& \rho k_{B}T+\frac{1}{3V}\left\langle 
\sum_{i<j}\textbf{f}\left( \textbf{r}_{ij}\right) 
\textbf{r}_{ij}\right\rangle ,
\label{eq:pres}
\end{eqnarray}

\noindent where $\textbf{r}_{ij}$ is the vector that it connects 
particle \textit{i} with particle \textit{j}, 
\textbf{f(r)} =-\textbf{$\nabla$U(r)}. 
The symbol $\left\langle  ...\right\rangle$ indicates ensemble average. 

The mobility of particles is 
evaluated by the mean square displacement, given by 
\begin{equation}
\left\langle \Delta r(\tau)^{2}\right\rangle = 
\left\langle\left[r(\tau_{0}+\tau) - r(\tau_{0}) \right]^2\right\rangle \; .
\label{eq:deslo}
\end{equation}

The diffusion coefficient is then obtained from the 
expression above by taking the  infinite time limit, namely
\begin{equation}
D = \lim_{\tau\rightarrow \infty}
\frac{\Delta r\left(\tau\right)^{2}}{6\tau}\; .
\label{eq:dif} 
\end{equation}

For normal fluids the diffusion at constant temperature grows with decreasing
density. Actually in most cases it is expected that it would follows 
the Stokes-Einstein relation, i.e., $D \propto T$.

The structure of the system studied by using the
translational order parameter, defined as~\cite{Sh02,Er01,Er03}
\begin{equation}
 t = \int_0^{\xi_c} \left| g(\xi) - 1 \right| d\xi,
 \label{eq_trans}
\end{equation}

\noindent where $\xi=r\rho^{1/3}$ is the inter-particle distance divided by the
average separation between pairs of particles 
$\rho^{-1/3}$. $g\left( \xi\right)$ is the distribution function 
of pairs. $\xi_{c}$ is the distance cutoff, where we use
half of the length of the simulation box, $r_{c}$, multiplied by 
$\rho^{1/3}$. Another alternative to $r_{c}$ would be the first 
or the second peak in the $g\left( r\right)$. Our choice is 
preferable, first, because it is the maximum distance allowed 
for the calculation of $g\left( r\right)$ \cite{Frenkel} giving 
us a better approach allowed for $t$. Second, the 
peaks of $g\left( r\right)$ change place according to 
density and temperature of the system. Thus
additional work would be necessary to find such positions.

For the ideal gas, $g=1$ thus $t=0$. As the 
system becomes more structured a long range order ($g\neq1$) appears 
and $t$ assumes large values. The translational order parameter has its 
maximum value in the crystal phase . Therefore, $t$ gives a measurement of
how close is the fluid close to the crystallization. 
For a fixed temperature normal fluids present
a monotonic $t(\rho)$ curve, increasing with density.

\section{Results} \label{results} 
\subsection*{Phase Diagram}
\begin{figure}[htb]
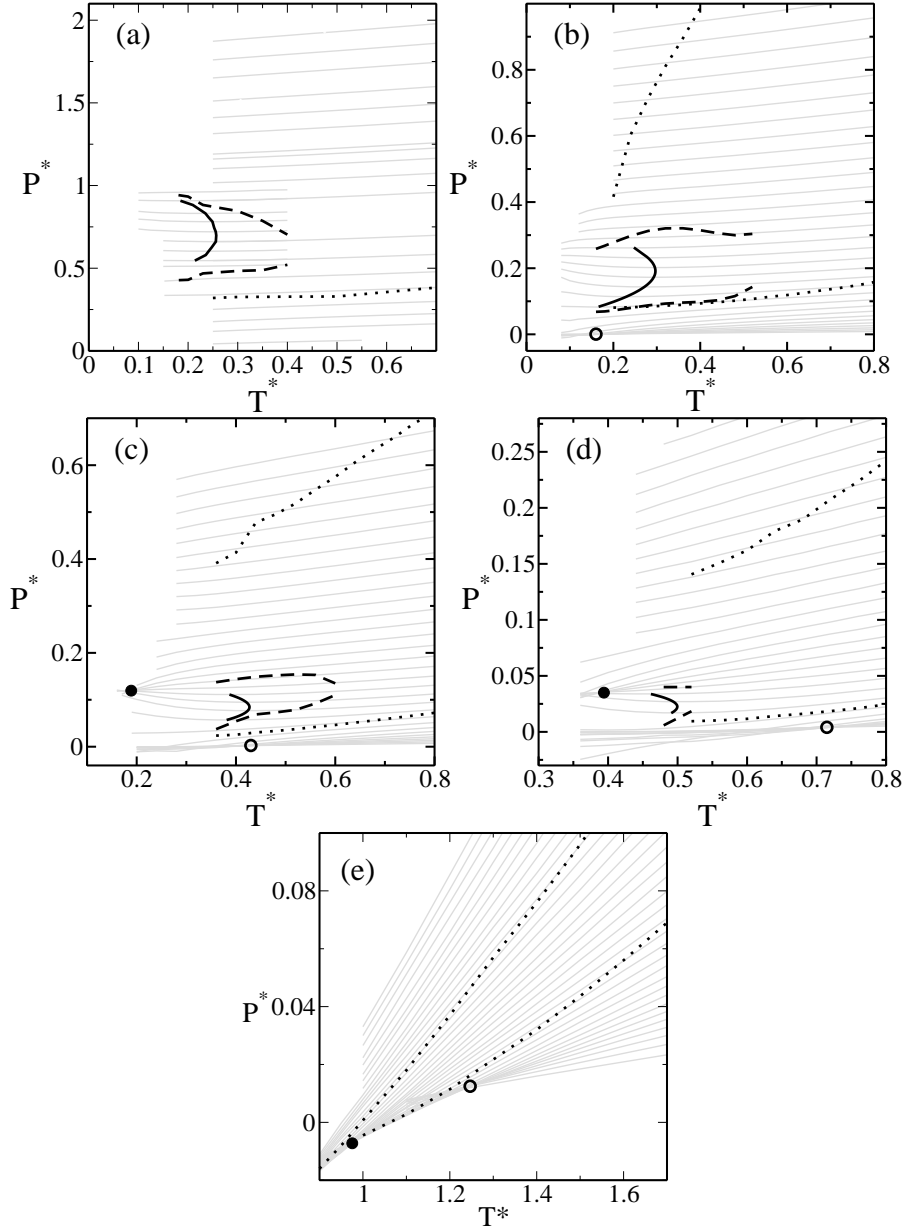

  \begin{centering}

  \includegraphics[clip=true,scale=0.3]{pt_m0.0.eps} 
\includegraphics[clip=true,scale=0.3]{pt_m0.25.eps}
  \includegraphics[clip=true,scale=0.3]{pt_m0.50.eps} 
\includegraphics[clip=true,scale=0.3]{pt_m0.75.eps}
 \includegraphics[clip=true,scale=0.3]{pt_m1.0.eps}
 
  \end{centering}
  \caption{Pressure-temperature phase diagram for the five cases studied 
in this work. The gray lines are the isochores. (a) Case  A ($b=0$): $\rho^*=$ 
0.04, 0.06,  0.07, 0.08, 0.09, 0.10, 0.107, 0.11, 0.115, 0.120, 0.125, 0.130, 0.134, 
0.140, 0.144, 0.148, 0.154, 0.158, 0.160, 0.168, 0.174, 0.180, 0.188, 0.194, and 0.20 from bottom to top. 
(b) Case B ($b=-0.25$):  $\rho^*=$
 0.01, 0.015, $\dots$, and 0.165 from bottom to top. (c) Case C ($b=-0.50$):
same as panel (b),
 (d) Case D ($b=-0.75$): same as panel (b). (e) Case E ($b=-1.00$): $\rho^{*} =$ 0.02, 0.025, $\dots$, and 0.2
 from bottom to top. The solid, bold line is the TMD line, the dashed line
  mark the maxima and minima in the diffusion and the dotted line bounds the region of structural anomaly. 
  The filled and open circles are the liquid-liquid  and liquid-gas critical points respectively.  }
  \label{fig:phase-diagrams} 
\end{figure}

Fig.~\ref{fig:phase-diagrams} illustrates the pressure-temperature
phase diagram for the cases A-E obtained through simulations using
the potential shown in Fig.~\ref{fig:potential}. As the attractive well becomes
deeper, the liquid-gas critical point appears and goes
 to higher temperatures
what can be easily understood as follows. At low densities the
liquid-gas transition is observed by cluster expansion namely
\begin{eqnarray}
\frac{\beta P}{\rho}&=& 1- 2\pi\rho\int_{}^{} 
f(r) r^2 dr  -\frac{8\pi^2\rho^2}{3} \int_{}^{} \int_{}^{}  
\int_{}^{} f(r) f(r') f(|r-r'|) \sin\theta r^2 r'^2 dr dr' d\theta 
\label{eq:pressure}
\end{eqnarray}
where $f(r)=e^{-\beta U(r)}-1$. The critical point is located at
\begin{eqnarray}
\frac{\partial P}{\partial \rho}&=&0 \nonumber \\
\frac{\partial^2 P}{\partial \rho^2}&=&0 \;\; .
\end{eqnarray}

\begin{figure}[htb]
  \begin{centering}

  \includegraphics[clip=true,scale=0.4]{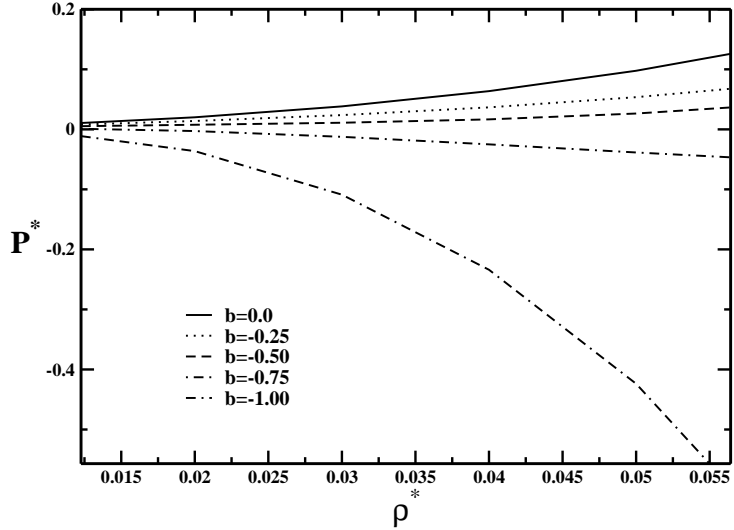} 

  \end{centering}
  \caption{Pressure versus density 
for $T^*=0.60$ for the cases $b=0.0,-0.25,-0.5,-0.75,-1.00$ from
top to bottom. }
  \label{fig:virial} 
\end{figure}

\begin{figure}[htb]
  \begin{centering}
   \includegraphics[clip=true,scale=0.4]{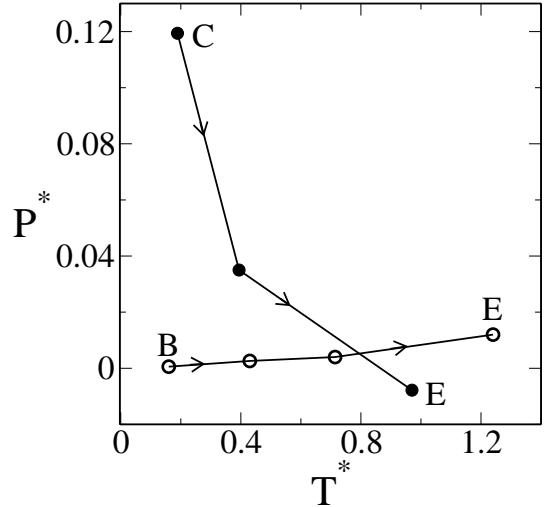} 
    \par
  \end{centering}
  \caption{Location of the critical points for the cases B-E considered in this work. 
  The Case A does not present any fluid-fluid critical point whereas the Case B has a liquid-gas but no liquid-liquid critical point.
  The symbols have the same meaning as in   Fig. \ref{fig:phase-diagrams},  i.e., filled and open circles 
mark the liquid-liquid  and liquid-gas critical points respectively. 
 The arrows indicate the direction of increasing the attractive interaction.
  }
  \label{fig:cp} 
\end{figure}

The low density behavior obtained
using the cluster expansion is
illustrated in Fig.~\ref{fig:virial}. For $T^*=0.60$ 
 Fig.~\ref{fig:virial} shows the pressure-density
phase diagram for $b=0.0,-0.25,-0.50,-0.75,-1.00$ using
the second and the third virial. For 
$b=-1.00$ the unstable region
of the pressure-density phase diagram 
is large and the system at this temperature is deep in the
liquid-gas coexistence region of the pressure-temperature
phase diagram. For $b=-0.75$ the unstable region is
present but is rather small. For $b=0.0,-0.25,$ and $-0.50$ no 
unstable region in the pressure-density
phase diagram is observed indicating that the system
is above the liquid-gas transition and that $T=0.60$ is 
larger than the critical point temperature. The comparison 
between the cases with
$b=0.0,-0.25,$ and $-0.50$ suggests that since the 
slope of the pressure-density
phase diagram increases as $b$ increases, the liquid-gas 
critical temperature
decreases as $b$ increases, 
$T_c^*(b=-0.25)<T_c^*(b=-0.50)<T_c^*(b=-0.75)< T_c^*(b=-1.00)$. Consequently
the attractive part favors the liquid phase to exists for higher
temperatures what is also observed in discontinuous
potentials \cite{Ma02,Ma05}.
Fig.~\ref{fig:cp} obtained from the simulations
illustrated in Fig.~\ref{fig:phase-diagram } summarizes the effect of the 
attractive part in the location of the critical points in the pressure-temperature diagram.

At high densities where the liquid-liquid phase transition is present
the cluster expansion with second and third virial is not appropriated.
Simulations show that as  $b$ decreases the pressure needed  to
form the high density liquid phase, decreases.
This reflects that the attractive part
favors the high density liquid phase over the low density liquid phase.
The attraction leads in this case to a more compact liquid phase
what is also observed in discontinuous potentials  \cite{Ma02,Ma05}.

\subsection*{Density anomaly}

The density anomalous region in the pressure
temperature phase-diagram can be found as follows.
From the Maxwell relation, 
\begin{equation}
 \left( \frac{\partial{V}}{\partial{T}}\right) _{P} = 
-\left( \frac{\partial{P}}{\partial{T}}\right) _{V}
\left(\frac{\partial{V}}{\partial{P}}\right) _{T},
\label{eq:TMD1}
\end{equation}
\noindent  the condition for density anomaly at fixed
pressure, i.e., a maximum in $\rho(T)$ curve 
given by $\left( \partial{\rho}/\partial{T} \right) _{P} = 0$, is 
equivalent
to the condition 
$\left( \partial{P}/\partial{T} \right) _{\rho} = 0$, corresponding
to a minimum in the $P(T)$ function.
While the former is suitable for $NPT$-constant experiments/simulations
the latter is more convenient for our $NVT$-ensemble study, thus 
adopted in this work.

In this sense, the minima at the  isochores mean that the system
has density anomaly. These extrema points in the pressure-temperature 
phase diagram
are named as temperature of maximum density (TMD) points, which connected form
the TMD line. Fig. \ref{fig:phase-diagrams}  shows the TMD 
line as a solid, bold line
in  panels (a)-(d), corresponding to the cases in which
$b=0.0,-0.25,-0.50,$ and $-0.75$ respectively. 
As the attractive well becomes deeper, the 
region in the pressure temperature phase-diagram occupied
by the density anomalous region  shrinks and moves to lower pressures
and higher temperatures until to the limiting case ($b=-1.00$)
in which no density anomaly is present [note that 
there are no local minima in the $P(T)$ 
curves of Fig.~\ref{fig:phase-diagrams}(e)].

\begin{figure}[htb]
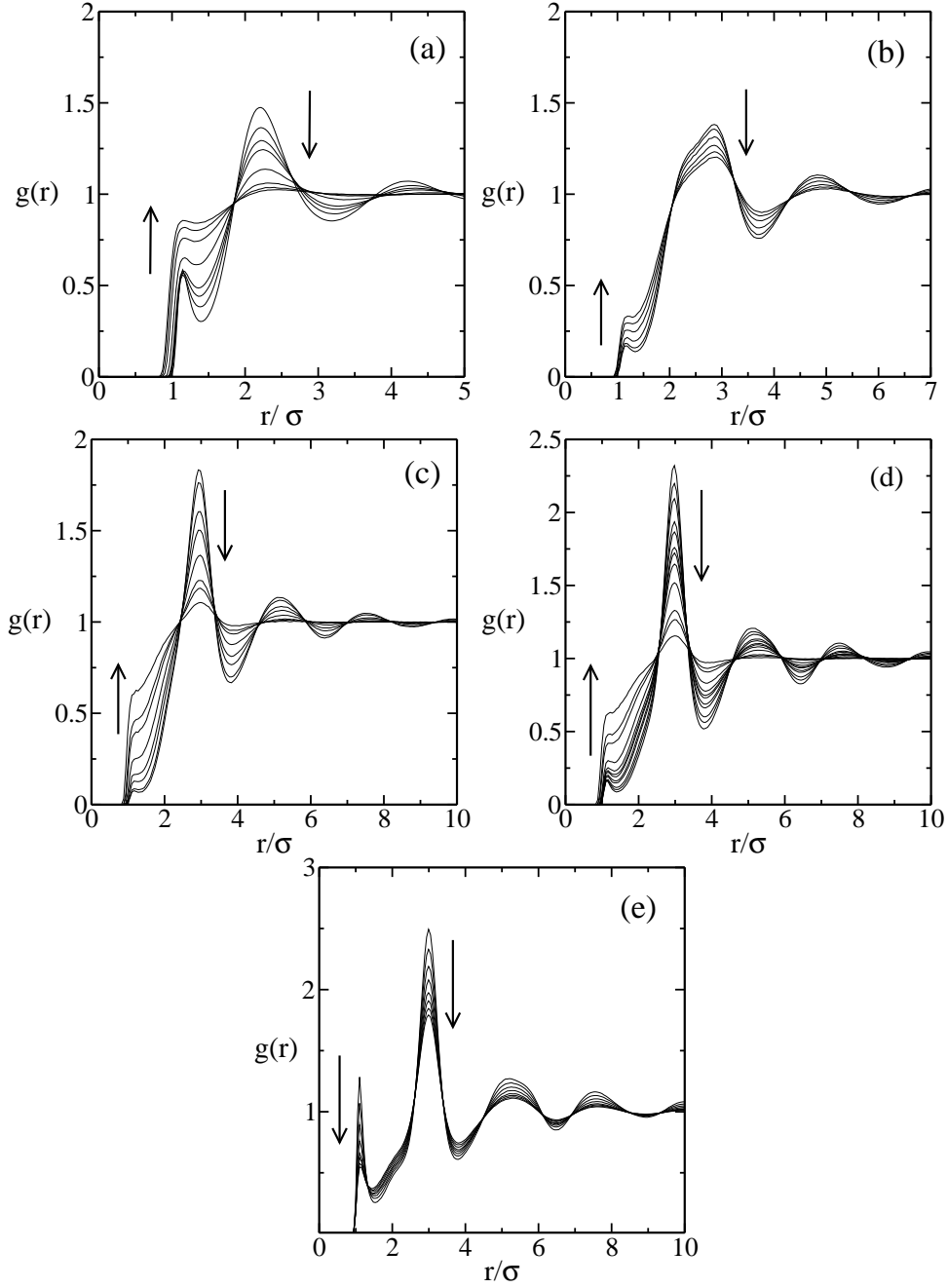

\begin{centering}

  \includegraphics[clip=true,scale=0.3]{gr-0.0.eps} 
\includegraphics[clip=true,scale=0.3]{gr-0.25.eps}
  \includegraphics[clip=true,scale=0.3]{gr-0.50.eps} 
\includegraphics[clip=true,scale=0.3]{gr-0.75.eps}
 \includegraphics[clip=true,scale=0.3]{gr-1.0.eps}
 
  \end{centering}
  \caption {Radial distribution  function versus distance for (a) the Case A ($b=0.0$) with
$\rho^*= 0.14$ and $T^{*}=0.25,0.35,0.45,0.55,1.0,2.0,3.0,$ and 4.0; 
(b) Case B ($b=-0.25$) with $\rho^*= 0.085$ and $T^{*}= 0.32,0.36,0.44,0.56,0.68,$ and 0.80; 
(c) Case C ($b=-0.50$) with $\rho^*= 0.06$ and $T^{*}=0.44,0.48,0.60,0.72,1.0,1.6,2.0,$ and 3.5;  
(d) Case D ($b=-0.75$) with $\rho^*= 0.06$ and $T^{*}=0.40,0.44,0.48,0.56,0.60,0.68,0.72,0.80,$ and 1.0; and 
(e) Case E ($b=-1.00$) with $\rho^*= 0.07$ and $T^{*} = 0.45,0.50,0.55,0.60,0.65,0.70,0.75,$ and 0.80. 
The arrows indicate the direction of increasing temperature.}
  \label{fig:gr} 
\end{figure}

The link between the depth of the attractive region and the presence or 
not of the TMD goes as follows. The TMD is related to
the presence of large regions in the system in which particles are 
in two preferential distances represented by the first scale and the 
second scale in our 
potential ~\cite{St98,St99,St00,Ol09}.  While 
for normal liquids as the temperature is increased
the percentage of particles at closest scales 
decreases [see the case $(e)$ in Fig.~\ref{fig:gr}], 
for anomalous liquids [see cases $(a)$, $(b)$, $(c)$ and $(d)$ in the 
Fig.~\ref{fig:gr}] there is a region 
in the pressure-temperature phase diagram where 
as the  temperature is increased the percentage 
of particles at the closest distance increases while
the percentage of particles in the second scale decreases. This
increasing in the percentage is only possible if particles
move from the second to the first scale. 

In Fig.~\ref{fig:gr}(e), the decrease of particles in the first scale
leads to a decrease of density with an increasing  temperature: behavior
expected for normal liquids. In the Fig.~\ref{fig:gr}(a)-(d), the 
increase of
particles in the first scale leads to an increase of density
with temperature what characterizes the anomalous region.
The anomaly is, therefore, related to the increase of 
the probability of particles to be in the first scale
when the temperature is increased while the percentage of 
particles in the second scale decreases.  As 
the potential becomes highly attractive this
``mobility'' between scales disappears, i.e., the
high density liquid becomes 
dominant and no anomalous region is observed.

\subsection*{Diffusion anomaly}

\begin{figure}[htb]
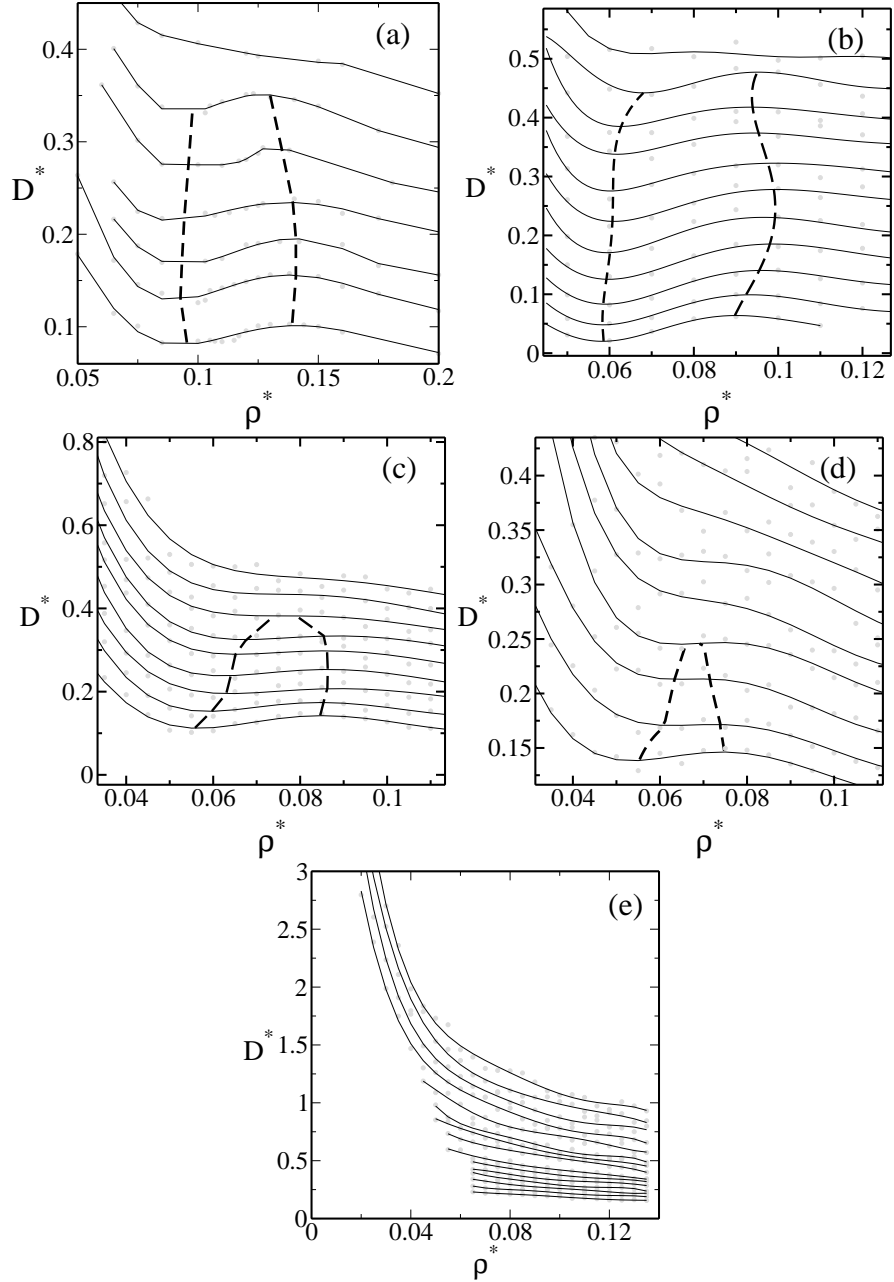

  \begin{centering}
  \includegraphics[clip=true,scale=0.3]{dif_vs_rho_m0.0.eps} 
\includegraphics[clip=true,scale=0.3]{dif_vs_rho_m0.25.eps}
  \includegraphics[clip=true,scale=0.3]{dif_vs_rho_m0.50.eps} 
\includegraphics[clip=true,scale=0.3]{dif_vs_rho_m0.75.eps}
\includegraphics[clip=true,scale=0.3]{dif_vs_rho_m1.0.eps}
    \par
  \end{centering}
  \caption{The diffusion coefficient against density for the (a) Case A, with 
isotherms 0.2, 0.23, 0.262,
  0.3, 0.35, 0.4, and 0.45 from bottom to top. (b) Case B with 
isotherms 0.16, 0.20, $\dots$, and 0.56, 
  (c) Case C, whose temperatures shown are 0.36, 0.40, $\dots$, and 
0.68, (d) case D, with temperatures
  0.48, 0.52, $\dots,$ and 0.80, and (e) Case E 
  with isotherms 0.70, 0.75, $\dots$, 1.0, 1.10, $\dots$,  and 1.70. 
  The dashed lines mark the local 
maxima/minima in the $D(\rho)$ curves. For the
 region enclosed by these lines particles move faster under 
compression. The dashed lines in this figure have
  the same meaning as those ones in Fig. \ref{fig:phase-diagrams}.}
  \label{Dif} 
\end{figure}

The mobility of any liquid is given 
by the diffusion constant. 
Figure~\ref{Dif} shows the behavior of the dimensionless translational 
diffusion coefficient, $D^{*}$, as function of the dimensionless 
density, $\rho^{*}$, at constant temperature 
for $b=0.0$, $-0.25,$ $-0.50,$ $-0.75,$ and $b=-1.00$. 
The solid lines are polynomial fits to the data obtained 
through simulation (dots in
Fig. ~\ref{Dif}). For normal liquids, the diffusion coefficient at constant 
temperature decreases with density. For the  cases A-D 
[shown in Fig.  \ref{Dif}(a)-(d)]  
$D^{*}$ anomalously increases with density in a certain range of 
pressures and temperatures instead. From Figure~\ref{Dif} 
we see that for very small and 
very high densities  $D^*$  decreases with increasing density as expected for
a normal liquid. For intermediate values of 
density, $\rho_{Dmax}>\rho>\rho_{Dmin}$,  $D^*$ increases 
with increasing density what leads to  local maxima at $\rho_{Dmax}$  
and a local
minima at $\rho_{Dmin}$. These local extrema in the diffusion versus 
density plots bound the 
region inside which the diffusion behaves anomalously (dashed lines 
in  Fig. \ref{Dif}). This region is mapped into the pressure-temperature 
diagram
illustrated in  Fig. \ref{fig:phase-diagrams} as dashed lines in  (a)-(d).
As the attractive well becomes deeper, the diffusion
anomalous region in the pressure-temperature phase diagram   shrinks and
it  goes to lower pressures. In the case in which $b=-1.00$, shown in  Figure~\ref{Dif}(e), the diffusion constant
behaves as in a normal liquid. This result again is consistent with the idea that a deeper attractive term favors 
the high density liquid phase.

\subsection*{Structural anomaly}

\begin{figure}[htb]
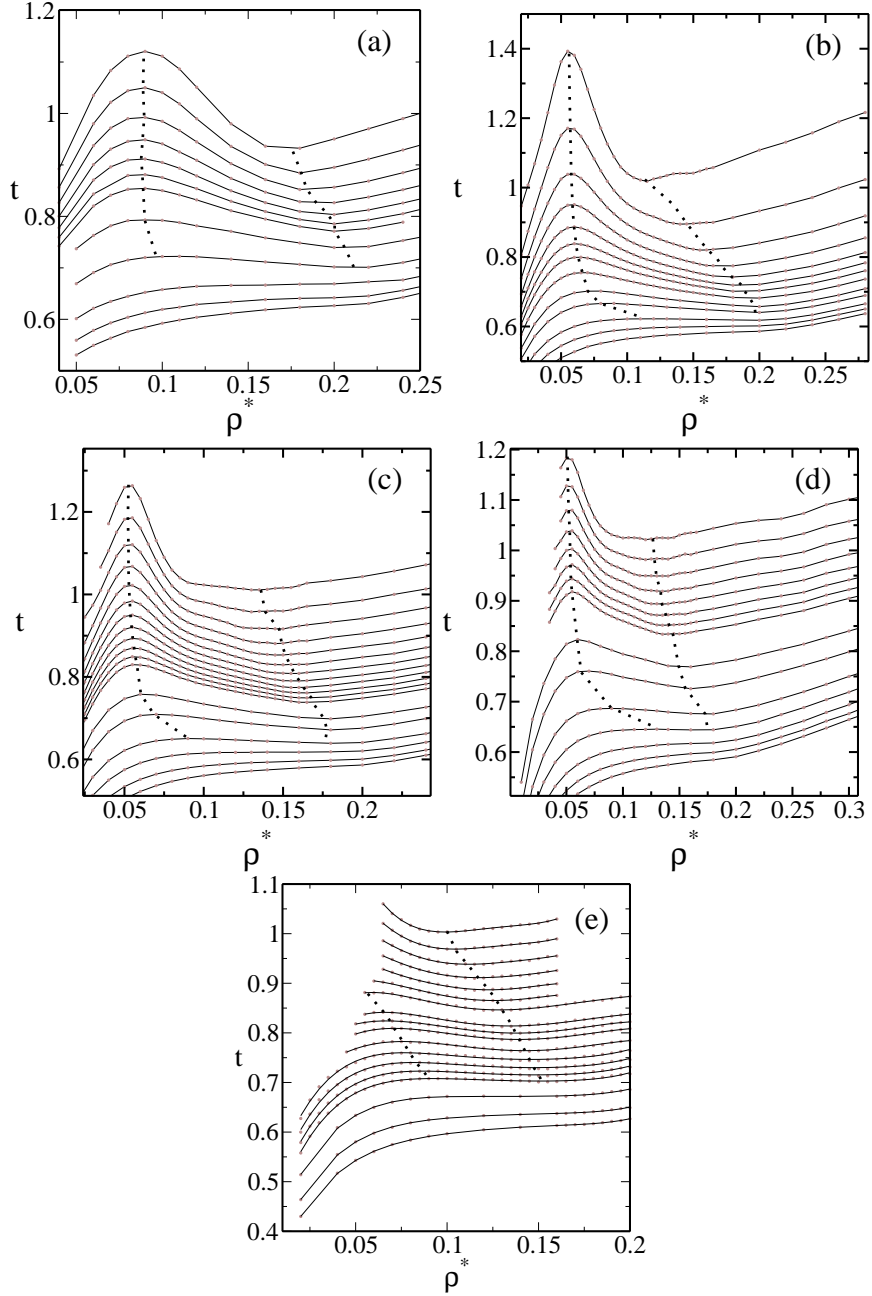

  \begin{centering}
  \includegraphics[clip=true,scale=0.3]{t_vs_rho_m0.0.eps} 
\includegraphics[clip=true,scale=0.3]{t_vs_rho_m0.25.eps}
  \includegraphics[clip=true,scale=0.3]{t_vs_rho_m0.50.eps} 
\includegraphics[clip=true,scale=0.3]{t_vs_rho_m0.75.eps}
\includegraphics[clip=true,scale=0.3]{t_vs_rho_m1.0.eps}
    \par
  \end{centering}
  \caption{Translational order parameter against density for (a) Case 
A, where each line
  correspond to an isotherm. The isotherms 
are: 0.25, 0.30, $\dots$, 0.55, 0.7, 1.0, 1.5, 2.0, and 2.5 from 
  top to bottom. (b) Case B, with isotherms 
0.20, 0.28, $\dots$, 0.68, 0.80, 1.0, 1.2, 1.6, 2.0, and 2.5 from 
  top to bottom. 
  (c) Case C whose temperatures are 
0.36, 0.40, $\dots$, 0.80, 1.0, 1.2, 1.6, 2.0, 2.5, and 3.0 from 
  top to bottom.
  (d) case D with $T^{*} = $ 0.52, 
0.56, $\dots,$ 0.80, 1.0, 1.2, 1.6, 2.0, 2.5, 3.0, and 3.5 from 
  top to bottom. Finally,
  (e) case E with $T^{*} = $ 0.70, 0.75, $\dots$, 1.0, 1.10, $\dots$,1.70, 2.0, 2.5, and 3.0 from top to bottom.
  The dotted lines bound the region of structural anomalies, i.e., the region where
  the parameter $t$ decreases upon increasing density.}
  \label{translat} 
\end{figure}

Besides the density and the diffusion anomalies an structural
anomalous region might be present.
Figure~\ref{translat} shows the translational order parameter as a 
function of density for fixed temperatures for
the potential we are studying for $b=0.0, -0.25, -0.50, -0.75,$ and $-1.00$.  
The dots represent the simulation data and the solid lines are polynomial fit to the data. 

The non-monotonic behaviour of these curves indicate
that there is a region in which $t$ decreases with density.
This means that the system becomes less structured 
for increasing density. Dotted lines determine the  
local maxima and minima of $t$, bounding the structural
anomalous region. This region was mapped into the pressure-temperature
phase diagram (dotted lines), as can be seen 
in Figure~\ref{fig:phase-diagrams}. The
comparison between the behavior for different $b$ values
indicates that as the attractive well becomes deeper the
structural anomalous region in the pressure-temperature
phase diagram shrinks and moves to lower pressures
and it is still present even in the deepest case, $b=-1.00$.
According to these results we believe that for $b<-1.00$, i.e.,
cases in which the attractive part is more intense than one showed in 
Case E, the structural anomalous region will also 
vanish. This result again is consistent
with the idea that a deeper attractive term favors 
the high density liquid phase.

Figure \ref{fig:anoms} gives an overview of the density, diffusion, and
structural anomaly locations in the pressure-temperature
phase diagram.

\section{Conclusions} 
\label{conc} 

\begin{figure}[htb]
 
   \includegraphics[clip=true,scale=0.35]{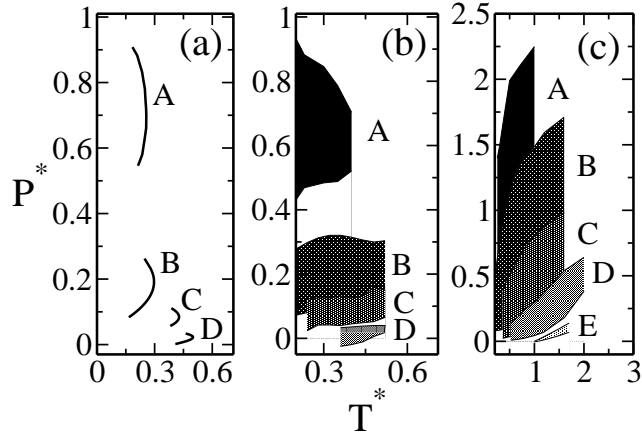}

  \caption{(a) TMD line for the cases considered in this paper. 
  Note that there is no density anomaly in the Case E (see Fig. \ref{fig:phase-diagrams}). 
  (b) The diffusion anomaly 
region for the Case A-D. No diffusion anomaly was found for the Case E (see fig. \ref{Dif}). The shadowed regions
  correspond to the region between the dashed lines 
in Fig. \ref{fig:phase-diagrams}. In (c) is shown the structural 
anomalous region for Case A-E. Here, the shadowed region corresponds to the region 
between the dotted lines in Fig. \ref{fig:phase-diagrams}.
See the text for discussion.}
  \label{fig:anoms}
\end{figure}

In this paper we have explored the effect of the addition
of an attractive part in a two length scales potential. Particularly
we analyze if the depth of the attractive part changes 
the position (and the presence of not)
in the pressure-temperature phase diagram of
the two liquid-gas and liquid-liquid critical points and 
of the density, diffusion and structural anomalous regions.

For sufficiently intense attraction between particles 
both the liquid-liquid and the liquid-gas critical points are present.
These two critical points are observed even for a very attractive potential.
For a small attractive interaction, only the liquid-gas critical point was 
found what indicates that for the coexistence of two liquid phases the
attractive well have to be deeper than a certain threshold.

Since the attraction favors the liquid phase (particularly the
high density liquid phase), as the  $b$ decreases  the liquid-gas critical point 
moves to higher temperatures (shown in Fig.~\ref{fig:cp}) and the 
liquid-liquid critical point to lower  pressures.

The density, diffusion and structural anomalous regions are
present  even in the absence of attraction. As $b$ decreases, the high
density liquid structure is favored and so the anomalous regions 
in the pressure-temperature phase diagram ( shown in 
Fig.~\ref{fig:anoms})
shrinks, moves to lower pressures and disappears for
very attractive potentials. 

In resume density and diffusion anomalous regions are present
in two length scales potential if the attractive interaction
is not  too strong.

\section*{ACKNOWLEDGMENTS}

We thank for financial support from the Brazilian science agencies CNPq, CAPES and FAPEMIG. 
This work is also partially supported by the CNPq through the INCT-FCx.

\bibliographystyle{aip}
\bibliography{Biblioteca}

\end{document}